# Title

Four Bottomless Errors and the Collapse of Statistical Fairness

# Author

James Brusseau
Philosophy and Computer Science, Pace University, New York City, USA
Department of Information Engineering and Computer Science. University of Trento, Italy

# Abstract

The AI ethics of statistical fairness is an error, the approach should be abandoned, and the accumulated academic work deleted. The argument proceeds by identifying four recurring mistakes within statistical fairness. One conflates fairness with equality, which confines thinking to similars being treated similarly. The second and third errors derive from a perspectival ethical view which functions by negating others and their viewpoints. The final mistake constrains fairness to work within predefined social groups instead of allowing unconstrained fairness to subsequently define group composition. From the nature of these misconceptions, the larger argument follows. Because the errors are integral to how statistical fairness works, attempting to resolve the difficulties only deepens them. Consequently, the errors cannot be corrected without undermining the larger project, and statistical fairness collapses from within. While the collapse ends a failure in ethics, it also provokes distinct possibilities for fairness, data, and algorithms. Quickly indicating some of these directions is a secondary aim of the paper, and one that aligns with what fairness has consistently meant and done since Aristotle.

# Keywords

Fairness, statistical fairness, AI, ethics, philosophy

# 1. Introduction

Statistical fairness is today's central approach to questions about the distribution of benefits and harms in AI ethics discussions when led by computer scientists – it is the way computer scientists *do* ethics (Carey and Wu 2023). It is also a particularly insidious kind of error. Statistical fairness is an error *of* reasoning, not *in* reasoning. Because erroneous fairness conceptions precede fairness thought, the result is a double predicament. Fairness outcomes are initially mistaken, and then inescapable since attempts to rectify only repeat the initial misstep causing the problem. Ultimately, the larger project collapses into its own futility.

Bottomless errors are not like 2 + 2 = 5, they are like 2 + 2 = apple. They are not wrong in some relative sense, but wrongheaded, which explains why modifying the answer in one direction or another – changing from apple to orange – gets you no closer and no further from the truth, it only echoes the initial mistake. Every answer becomes a mistake, even before it is proposed. This essay traces four examples of this futility, and the conclusion will not be that promoters of statistical approaches have been miscalibrating their fairness implementations or off-target in their understandings. The conclusion is that there is no incremental solution. Statistical fairness advocates should stop working and delete what they have done.

Then the way will open to a meaningful connection between fairness and statistics and artificial intelligence, one that returns to the basics by grasping the meaning of fairness first, and by applying it algorithmically only subsequently. In this paper, some indications of that return will be developed, but the primary purpose is to do justice to the title by collapsing statistical fairness as it exists today so that the approach will be abandoned, and the work of ethics can recommence on a level that intersects with AI.

## 2. What is statistical fairness?

Statistical fairness can be defined superficially as a set of authors, articles, and conferences where the work appears. Many of the representative figures and documents are gathered in this paper, and together they form a movement, a common set of ideas and practices. More penetratingly, statistical fairness is a method. It applies statistics to produce conceptions of fairness, as opposed to starting with a conception of fairness and then applying it statistically.

The reasoning sequence can be directly experienced in the conference presentation *Individual Statistical Fairness* where the presenter relates that, "we've come to realize during this workshop that definitions are tricky. We can say that something was unfair, but we have to try to be precise about what we mean (Roth 2019: 01:40)." The reasonable assumption is that a precise definition will follow. The presenter will say, "Unfairness is…." Then the definition can be justified with arguments, and subsequently applied with mathematical tools. Instead, what follows is a tour of statistical distributions of possible college admissions formulas. The evasion persists until the realization dawns that the distributing *is* the response. The algorithmic arrangements *are* what is meant by fairness. Math creates ethics, instead of ethics determining the uses of math.

The method is less explicit but equally pervasive in the seminal paper *Fairness Definitions Explained* (Verma and Rubin 2018), and in the milestone conference presentation *21 Fairness Definitions and their Politics* (Narayanan 2018), and in the notorious debate between COMPAS and ProPublica (Angwin et al. 2016. Rudin et al. 2020). What holds the examples together is that statistical processing generates an ethical principle, as opposed to an explicit principle guiding how information is processed.

The method is a vulnerability. The reason statistical approaches go so far wrong – the reason they are not even on the continuum between right and wrong – is that the primary question *What is fairness?* is miscategorized as a mathematical as opposed to an ethical entity. It is inversely analogous to using ethics to do math, it would resemble saying that the rules of multiplication must be adjusted because they are unfair to prime numbers since the primes receive fewer factors than the others. The result would be factors added in the name of equality, so 2 x 11 and 3 x 8 would both equal 23 according to this moralized math, which is bazar, and also an illuminating mirroring of what statistical fairness does on the ethical level.

## 3. What is fairness?

Fairness has been stably defined for nearly 2500 years: Treat equals equally and unequals proportionately unequally (Aristotle 350 BC: Book 5, 3A).

While peripheral debates about the conception trace through philosophy's history (Broome 1990), at least two central claims remain solid. First, fairness is about a process and not an outcome. This is why a coin flip is fair. While it results in unbalanced treatment, it remains justifiable since equals are still treated equally in the sense that each participant had the same chance in the process (Redacted 2024). Second, fairness is about individuals before society. Treating individuals fairly, in other words, results in an ethically justifiable society, as opposed to the idea that an ethically justifiable society will dictate what fairness means for individuals. This is the core of Nozick's (1974) Wilt Chamberlain thought experiment and argument against Rawls's political philosophy. The experiment explains why Chamberlain's exceptional talent should be rewarded unequally as compared with other players, and shows why the resulting society is justified because the treatment of its individuals is fair.

These details about fairness's definition threaten to be distractions, however, from the main point of this paper. What is significant is not what Aristotle's definition means, instead, it is that the definition leads to debates about fairness within the parameters of ethics and reason. What this essay shows is that statistical approaches do not exist within those parameters. They have no relation – neither positive nor negative – with the ethics of fairness. So, it is not that Aristotle's approach is better ethics than statistical fairness approaches, it is that Aristotle is ethics and statistical approaches are not.

## 4. Equality error

Fairness can be misconceived as straight equality – everyone treated the same – as in this example:

> Fairness is the absence of any prejudice or favoritism toward an individual or group based on their inherent or acquired characteristics.

This particular rejection of prejudice and favoritism creates an insuperable problem: no discriminating decisions can be made. There is nothing to judge as fair or unfair. A basketball team cannot seek tall (or short) players because that favors an inherent characteristic. Universities cannot reject (or accept) applicants with abysmal grades since that exhibits prejudice against an acquired quality. Once the pattern is established, the reduction to absurdity speeds to its conclusion: an ethical tool ostensibly producing thoughtful decisions about the distribution of opportunities ends up prohibiting any thought at all about justifying distributive decisions. It is not that the definition is wrong, it is that it refutes the condition of its own possibility. The idea of fairness deteriorates into parody.

Of course, it is always easy to invent an internally contradictory concept and then attack it with flourish, but there are no strawmen here. The quoted fairness definition comes from a real publication, *Computing Surveys*, which carries an elevated impact factor and more than fifty years of publication. More, the specific article, *A Survey on Bias and Fairness in Machine Learning* (Mehrabi et al. 2021), has been cited more than three thousand times, with authors ranging from academic computer science (Morse et al. 2022) to medical researchers employing AI (Benjamin et al. 2024) to private industry (Varona and Suárez. 2022).

So, the subject here is not a flawed detail infiltrating a few marginal publications. It is mainstream thought impacting the academics of computer science and the practical world of information engineering where fairness decisions are being made every day in critical environments including hospital emergency rooms. The fairness definition, in other words, is widely respected and socially determinative. And, across the entire range it is also a bottomless error because a misunderstanding about fairness is prohibiting the kind of discriminating answers that justify questions about fairness in the first place.

Then things get worse. Not only does the definition collapse, it also twists back to expand the error. After misrepresenting fairness as equality, there follows the wider mistake of imagining that where there is equality, there is also fairness.

Equality as fairness is rampant in today's AI ethics. One of the more influential examples is *Fairness Definitions Explained* (Verma and Rubin 2018) where the error lies not in the explanations of the plural definitions of fairness, but in *how* the plurality occurs. It occurs by multiplying forms of equality: statistical parity, predictive parity, false positive error rate balance, equalized odds, overall accuracy equality. Every one of these conceptions generates from an ethical logic moving in the wrong direction. "Parity" "balance" and "equality" convey the reasoning that *because* there is an equals sign, fairness is in effect.

The effect has not gone unnoticed in the community: "In the data science world, when they try to put down what fairness means, it inevitably becomes something is equal to something" (Meng and Vittert 2022: 13:15) As for *what* is equal, that reduces to a detail within the controlling logic. And as for that controlling logic, it cannot be appealed because it is inherent in how statistical fairness *begins*: the symbol, signs, and logic of mathematics precedes ethical considerations, statistics dictate fairness instead of fairness arranging statistics.

Regardless of the cause, the consequence is circling, inescapable error: fairness means equality, and equality means fairness.

### 4.1 How does the equality error provoke future algorithmic fairness?

One provocation of fairness misunderstood as equality is the more accurate conception of fairness as primarily – though not only – about treating people *un*equally. In hiring, someone gets the job while others do not. So too in university admissions and most places where fairness dilemmas occur: ethics is about differences.

For that reason, the primary fairness and statistical challenge is how to distribute uneven treatment, in what ways, and according to which metrics? For practical work in artificial intelligence ethics, these are the core fairness questions, they are about *un*balancing, not balancing. One area where work has been done on this subject is AI-driven image analysis for skin cancer (Aggarwal 2021) where skin tones are unequally vulnerable and therefore require proportionately unequal treatment. (Redacted 2021, Redacted et al. 2021).

### 5. Perspective error

*21 Fairness Definitions and Their Politics* is a standard reference in statistical fairness (Castelnovo et al. 2022, Courtland 2018), and the presentation's central use case is the COMPAS algorithmic crime predictor, which estimates whether an arrested defendant will reoffend if released from jail while awaiting trial (Northpointe 2015). Essentially, the prediction is low or high risk, which translates into eligible or ineligible for bail.

Regarding the twenty-one fairness definitions, they are attributable to distinct individual contexts (Narayanan 2018: 13:15, 21:45) and, according to one of them, fairness increases as false positive recidivism predictions drop toward zero (Narayanan 2018: 14:05). Everyone jailed, in other words, *would* have gone on to commit another crime had they been freed. Stated positively, the goal is for no one to be imprisoned unnecessarily and, in the courtroom, the human result is a formula that bails out close calls. Intuitively, there is a sense of fairness here (Hardt 2020: 01:21:10), and also one that appeals to defendants as they avoid needless prison time (Narayanan 2018: 13:20).

Something is left out, though, an entire side of the experience. If the classification error is tilted to release borderline cases, then some of those are going to commit more crimes. That means new victims will be created, and from *their* perspective – from their suffering robbery, rape, assault – the same bias that tilted toward release will flip: for them, the intuitive statistics of fairness will drive toward zero the number of defendants incorrectly classified as low risk. Instead of no unnecessary jail victims, there are no unnecessary crime victims.

What emerges are conflicting viewpoints representable as recidivism predictions that are truly and falsely positive, and truly and falsely negative: TP/(TP+FP), TN/(TN + FN). The formulas are mathematical, ethical, and illuminating because instead of being an objection to statistical fairness, their incompatibility is an expectation (Rudin et al. 2020: Introduction). It is even the essence. It reveals *how* the theory that there are twenty-one fairness definitions works from its inception. It works by slicing distinct population segments and asking what fairness means for them myopically. This is critical: it is not that one perspective exists *among* others, but by *disregarding* others. The ethics, in other words, mimics the math where it is possible to round close calls down or up, but one will always be achieved by negating the other, by denying the other.

In the presentation, the speaker was specifically interested in articulating the defendants' perspective (Narayanan 2018: 13:20), but the significance is the commitment to fairness as *constituted* by discounting the experiences and understandings of others. It is not just that seeing the defendants' perspective means blindness to the victims, it is that blindness is the *way* the perspective exists in the first place. Exclusion as ethics is how statistical fairness comes into being: it multiplies some human beings by zero in the mathematical sense, and then transfers the reasoning up to the ethical level in order to create one of the presentation's 21 definitions.

When that happens, statistical fairness stops making sense. A concept deployed to help people settle differences together ends up eliminating some altogether. The solution contradicts the problem eliciting the solution. There is no bottom to this kind of structural error, though that does not render it insubstantial or confusing. The ethical logic of multiplying some people's perspectives by zero is palpable and easily recognizable behind the 20th century's most infamous horrors.

### 5.1 How does the perspective error provoke future algorithmic fairness?

The first question posed by any perspectival stance gathering multiple definitions of fairness is: How did the presenter know what to gather? On what grounds did he distinguish the twenty-one legitimate fairness formulas from others deemed unfair, or unrelated to fairness? The more general question is: If you do not know what something is, how will you know when you have found it? This was Plato's question at philosophy's beginning, and the answer has not changed: in some way, we must *already* have the idea. To find twenty-one definitions of fairness, you must already know the *one* definition that divides those formulas from the others that are not fair.

The work of ethics – for philosophers and for computer scientists – lies in investigating and articulating that implicit idea, the one the presenter had even before he began his talk and, more consequentially, even before he began doing math.

### 6. Disproportion error

Disproportion is a subset of the perspective error. Besides participating in the logic of exclusion as ethics, it positively leverages suppressed information in a dataset to generate misinformation.

The paradigmatic example is ProPublica's activist research publication dedicated to the bail or jail COMPAS algorithm. About it, ProPublica related:

> Our analysis found that black defendants who did not recidivate over a two-year period were nearly twice as likely to be misclassified as higher risk compared to their white counterparts (Angwin 2016).

The accusation here is racism because blacks labeled high-risk are twice as likely as whites to be incorrectly classified and therefore unnecessarily jailed. The authors were correct about the numbers, but neglected to mention other numbers: black defendants rated high-risk were also twice as likely to be *correctly* classified compared to their white counterparts, and so rightfully jailed.

Why was COMPAS twice as wrong *and* twice as right for these black defendants? Because there were twice as many in the high-risk defendant population (Redacted 2021, Redacted et al. 2021). In this context, an algorithm that predicts fairly will naturally accumulate more hits, and more misses, for the proportionately larger group. This is easy to observe in Figure 1 which documents how members of different races – equals – are treated equally by the justice algorithm when it comes to right and wrong decisions about going to jail or going home.

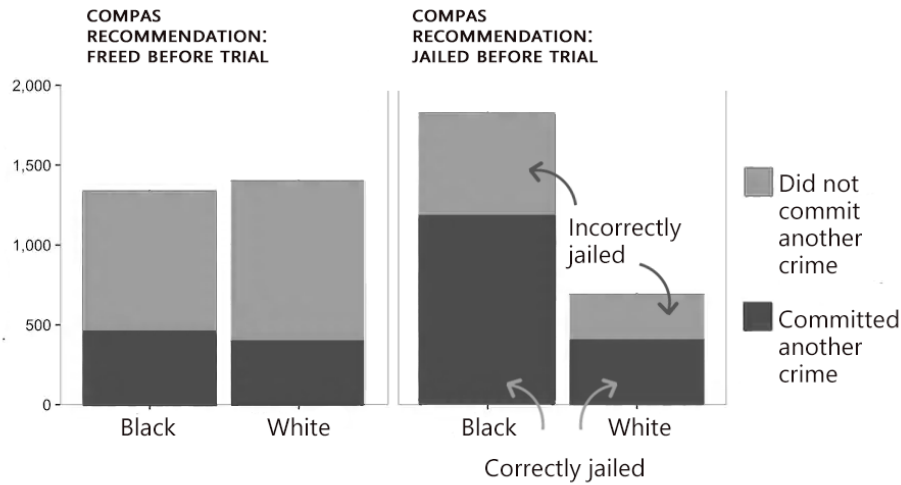

**Figure 1.** These tables are conceptually correct, though the specific numbers are in some dispute (Barenstein 2019).

For statisticians, there are overlaps here with Simpson's paradox and with the complexities of divergent base rates, but the technical details are less important than the wide potential for error arising when proportions are partially suppressed, when one side of an imbalance is highlighted while the other is understated, if not concealed. This is a sleight-of-hand trick, one that misleads with equal facility whether the fairness line is drawn through race, gender, age, language, religion, or any other personal characteristic.

For instance, there are many more male than female applicants for most data engineering roles, and naturally this will lead to more men than women being declined for in-person interviews. By the ProPublica logic, and by simply switching "wrongfully denied bail" for "wrongfully denied interview," in Figure 1, the resulting finding will be unfairness in computer science hiring…against men.

Regardless, what is significant is that when errors of proportion are spelled out in everyday language and clearly represented graphically instead of hidden behind the obscure jargon of AI statistical ethics – "positive predictive value," "false negative error rate balance" – the misinformation leaps off the page.

Despite the conspicuousness, the disproportion error keeps getting repeated, dismayingly even in publications and at places dedicated to expertise in just these matters (Pastaltzidis et al. 2022, Cole et al. 2022, Baumann et al 2022, Ganguli 2022, Bao et al 2021, Fogliato et al. 2021, Krafft et al. 2021, Fleisher 2021, Hao and Stray 2019). In any case, everywhere the disproportion error occurs there is futility echoing the perspectivism error: fairness claims are staked on radical unfairness – on the absence of fairness itself – by denying the ethical existence of those caught on one or the other side of an imbalance.

### 6.1 How does the disproportion error provoke future algorithmic fairness?

If the error is disproportion – a difference in how many subjects are compared with how many others – one remedy is comparison reduction to the level where proportions disappear: the individual.

Instead of measuring algorithmic fairness between groups, testing proceeds from single people and by comparing every one against every other. This is just the sort of numbingly repetitive task at which machines excel, and then, with the statistical valences accumulated, the set can be filtered for incongruencies and outliers. Pattern-breakers are critical because fairness is not about accuracy, only consistency. Even while it may be that increasing accuracy is the most practical way to solve many algorithmic fairness wrongs, it is certain that an algorithm that is always wrong can be perfectly fair. For the statisticians of fairness today, that means, efforts should be directed away from group considerations which potentially conceal numerical disproportions. What matters is congruence and coherency across the universe of individual data subjects.

### 7. Group fairness error

Because of the intersection with significant legal disputes and broad social justice, research in group fairness may be the most prominently debated area of statistical fairness (Mashiat et al. 2022, Chouldechova and Roth 2020, Binns 2020, Bird et al. 2020, Wachter et al. 2020). What is certain is that algorithms can be unfairly biased against social groups.

Something near the reverse is also true: fairness *is* the deployment of bias to create groups. An algorithm trained for college admission will likely link strong high school grades with college success, and then deploy a corresponding admission preference. Unequals (in grades) are treated unequally (in admission), meaning that fairly distributing opportunities generates social classes (admitted and rejected). Subsequently, these groups may be refined with labelling distinguishing female from male, white from black, and similar.

Complementarily, the group fairness error is *starting* with groups (black and white college applicants, for example, or male and female medical patients), and subsequently evaluating their treatment as fair or unfair (Wallach and Dudik 2021: 2:15; Andrus 2021: 00:30; Binns 2020: 0:55). These groups may be delineated by legal dictates, financial differences, gender, race, or other factors, but what matters is the concept: doing ethics means fairness produces groups, as opposed to having groups subsequently judged as fair or unfair.

One richly human scene of the error is the conference presentation *Inherent Tradeoffs in Algorithmic Fairness*. The talk presupposes the established groups of women and men, and progresses into a hypothetical dilemma between unequal medical risks arising because one group is more vulnerable than the other (Kleinberg 2018: 00:25:50), as occurs, for example, with breast cancer which men can contract, though far less frequently than women (CDC 2021). The problem is that algorithmic predictions are going to skew with the uneven vulnerabilities. If women are a hundred times more likely to suffer the cancer, it follows that an overall population composed about evenly between the genders will produce many more incorrect positive predictions for women than for men. Within the myopic world of statistical ethics, this can seem unfair: women suffer from misdiagnosis and the accompanying psychological anguish at a far higher rate than their gender contraries.

There are two solutions. One is to reduce the disproportionately high number of women suffering misdiagnosis. However, since the cancer-probability accuracy is presumably maximized, it is impossible to reduce female false positives without also losing true positives, and so creating even more imbalanced suffering. This leaves the other option for equalizing anxiety between genders: *Inject error into the predictions for men*. "It would mean," as the presenter describes, "that you would go to the doctor, and the doctor would say, 'Well we looked at some features and this thing reports a certain percent chance you will have the disease, but because you are male, we will have to adjust that…'" (Kleinberg 2018: 00:26:50) Specifically, it would need to be adjusted upward so that both women and men suffer misdiagnosis equally frequently.

This is a fascinating visual moment in the history of statistical fairness. In the video of the presentation, the speaker clearly grasps that his approach is internally conflicted. The cognitive dissonance wrinkles his face as his mind balances the irrefutable logic of his own fairness method against the disconcerting reality of doctors intentionally misdiagnosing patients in the name of ethics. Rationality coexists with absurdity, and in that moment the way opens to confront the full truth about statistical fairness. Not just this one example or even this one kind of error, but all the errors and the entire bulwark now verging on collapse under the weight of nothing more than its own conclusions. The end of statistical fairness is happening. He just needs to say it.

Instead, he recoils and veers toward a technical explanation of statistical calibration.

The mathematical deviation allows momentary alleviation from this truth: he was wrong about fairness before he began. When group fairness is conceived within the assumption of preestablished communities (men and women), and when the subject population is a typically human mix of characteristics spread irregularly throughout (breast cancer), the data and algorithmic functioning will twist into some version of unfairness (men and women suffer unequally from misdiagnoses). Next, if the attempt is made to fix the problem by changing the way the groups are treated, instead of changing the definitions of the groups that are treated so that they correspond with the data determining the treatment, then the mistake initially causing the problem repeats. The unfairness begot by thinking in terms of groups is addressed by again thinking in terms of groups. There is no escape because getting into the imbalance and trying to resolve it requires already subscribing to the error that generated the disequilibrium in the first place.

This plunging futility is what it looks like when *ways* of thinking collapse.

### 7.1 How does the group error provoke future algorithmic fairness?

Ethics and law are different kinds of reasoning – ethics hierarchizes human values while law articulates social dictates – but legal outcomes nevertheless get conflated with ethical evaluations (Hardt 2020: 00:14:55, Romano et al. 2020: Introduction, Watkins et al. 2022, Deho et al. 2022, Carey and Wu 2022, Zhou et al. 2022, Feldman and Peake 2021, Feuerriegel et al. 2020). One blunt case involves a presumed contradiction:

> If a higher proportion of applicants from group B are accepted than applicants from group A, the system might seem unfairly biased against group B (where A and B are different values of a protected characteristic). On the other hand, if two applicants who have similar SAT scores, extra-curricular activities, and interview scores are given different outcomes, the system may seem unfair in a different way (Binns 2020).

In fact, this is not a conflict between two kinds of unfairness, instead, it is between wrong and right understandings of how fairness relates to groups: either a group predefined by legal reasoning is taken as the subject of ethics, or ethics forms groups which later may be tested for legal validation.

The validation may be received: groupings initially based on fairness may flow in humanly-intuitive directions and then harmonize with legal frameworks. High grades may correlate with college success and subsequently delineate a legally sanctioned group favored for admission. But, because algorithms produce knowledge differently from humans – correspondence versus cause and effect – it also seems likely that some groupings will not be human-intuitive, or even be counter-intuitive (Redacted 2022).

Pushing the ethical and epistemic point, if unconstrained fairness is allowed to gather and distribute its subjects according to its own logic and powers, will startling groupings ensue? Will some feel chaotic, or incomprehensible, or maddening, or illegal? Historically, advances in knowledge have created just these rippling crosscurrents. The unexpected outrages are even part of how we realize we have advanced: Socrates, Copernicus, Galileo, Darwin. Perhaps we will know that AI fairness is actually happening when people start getting exiled for their algorithms.

**Disclosure Statement**

The author has no financial or non-financial disclosures to share for this article.

**References pending final inclusion/exclusion, formatting**

**End**